\documentclass[sigconf]{acmart}

\renewcommand\footnotetextcopyrightpermission[1]{} 
\setcopyright{none} 

\usepackage{comment}
\usepackage{endnotes}
\usepackage{xspace}
\usepackage{url} 

\usepackage{subfig}
\usepackage{color}
\usepackage{multirow}
\usepackage{enumitem}
\usepackage{wrapfig}
\usepackage{float}
\usepackage{graphicx}

\usepackage{algorithm,algpseudocode}

\usepackage{refcount}

\usepackage{lipsum}
\usepackage{mwe}

\newcommand{\FACADE}{{PASSAT}\xspace}

\begin{document}
%
\title{\FACADE: Single Password Authenticated Secret-Shared \\ Intrusion-Tolerant Storage with Server Transparency}

\author{Kiavash Satvat$^*$, Maliheh Shirvanian$^{\mathsection}$, Nitesh Saxena$^+$ }
\affiliation{%
	\institution{University of Illinois at Chicago$^*$ Visa Research Lab$^\mathsection$ University of Alabama at Birmingham $^+$}
}
\affiliation{%
	\institution{ksatva2@uic.edu$^*$, mshirvan@visa.com$^\mathsection$, saxena@uab.com$^+$}
}

\begin{abstract}
	In this paper, we introduce \FACADE, a practical system to boost the security
	assurance delivered by the current cloud architecture without requiring
	any changes or cooperation from the cloud service providers. \FACADE is
	an application transparent to the cloud servers that allows users to
	securely and efficiently store and access their files stored on public
	cloud storage based on a single master password.  Using a fast and
	light-weight \textit{XOR secret sharing} scheme, \FACADE secret-shares
	users' files and distributes them among $n$ publicly available cloud
	platforms. To access the files, \FACADE communicates with any $k$ out
	of $n$ cloud platforms to receive the shares and runs a secret-sharing
	reconstruction algorithm to recover the files.  An attacker (insider or
	outsider) who compromises or colludes with less than $k$ platforms
	cannot learn the user's files or modify the files stealthily.  To
	authenticate the user to multiple cloud platforms, \FACADE crucially
	stores the authentication credentials, specific to each platform on a
	\textit{password manager}, protected under the user's master password.
	Upon requesting access to files, the user enters the password to unlock
	the vault and fetches the authentication tokens using which \FACADE can
	interact with cloud storage. 
	Our instantiation of \FACADE based on \textit{(2, 3)-XOR secret sharing}
	of Kurihara et al., implemented with three popular storage providers, namely, \textit{Google
		Drive}, \textit{Box}, and \textit{Dropbox}, confirms that our approach can efficiently enhance the
	confidentiality, integrity, and availability of the stored files with
	no changes on the servers. 

\end{abstract}

\settopmatter{printacmref=false, printccs=true, printfolios=true} 

\thispagestyle{empty}
\fancyhf{} 

 	\maketitle

	\section{Introduction}
\label{sec:intro}

Cloud storage services are widespread and popular online services used
for archiving, backup, and data sharing.
A variety of service providers, including Amazon Web Services, Dropbox, Microsoft Azure,
Box, and Google Drive, offer cloud storage and file synchronization
 through application-based (and browser-based) interfaces.  
Despite the evident advantages of cloud storage, security and privacy concerns
refrain customers from full adoption of the cloud services
\cite{slamanig2012cloud,nocloud}. The current implementation 
demands customers' trust on  providers, whereby users cannot 
monitor and assure the security (including confidentiality and integrity) of their data.

Various approaches were proposed  to enhance the
security of cloud storage through encryption, replication, erasure code, and
secret-sharing \cite{bessani2013depsky,bowers2009hail,
bagherzandi2011password,stefanov2012iris,wilcox2008tahoe,resch,popa2011enabling,jarecki2014round}.  
However, given the complication of their use by non-expert users, key
management concerns, encryption bans in some countries, and  additional
computation/communication overhead they may pose in practice, the
adoption of them still remains questionable.  Critically, many of
these approaches require changes on the side of the public cloud service
providers, which may not be possible unless these providers cooperate, and 
therefore, most of these approaches cannot be currently deployed from the
user-side.


We propose \FACADE, an efficient password-controlled storage
paradigm to address the security concerns in storing users' private data on
untrusted platforms, without any need for modifications or cooperation from
currently available platforms.  \FACADE employs the notion of \textit{secret
sharing} to improve the confidentiality, integrity, and availability of the stored data. 
In the \FACADE protocol, the user has some private data $sc$  that she wants to
store and protect, and be able to later access on the basis of a single
password $pw$.  \FACADE employs a fast and light-weight XOR secret sharing
scheme to distribute the file ($sc$) into $n$ multiple distinct shares and
stores these shares across public cloud storage platforms.  To access the
files, \FACADE communicates with  any $k$ out of $n$ cloud platforms on the
input $pw$ from the user and runs a secret-sharing reconstruction procedure to
recover the file.  An attacker should compromise or collude with at least $k$
platforms to learn the user's files. Said differently, an attacker who
compromises or colludes with up to $k-1$ services and learns the corresponding
up to $k-1$ shares of the files cannot learn anything, in the information
theoretic sense, about the files themselves.  This enhances the security of
$sc$ against not only the malicious outsider attacks but also the insider
attacks (it is known that the cloud service providers may peek into user data
\cite{peek,pirate}). Moreover, none of the existing cloud providers are HIPAA
compliant, and there is no certification recognized by the US Department of
Health and Human Services for HIPAA compliance. Complying with HIPAA
is a shared responsibility between the customer and cloud services, which can
be addressed using \FACADE.   

In addition to improving the confidentiality and privacy of the stored data,
\FACADE enhances its integrity because any malicious modification of less than
the $k$ stored shares can be detected by the user during the reconstruction
process as such reconstruction will yield random garbage or meaningless data.
Protecting the integrity of the stored data is important
as a malicious or compromised cloud provider is able to
modify the data contents.
Furthermore, \FACADE improves the availability of the data and protects it
against deletion by the service providers (as previously reported in
\cite{deleted}) because the data remains available as long as $k$ valid shares
are accessible.

To access the information stored in the cloud, \FACADE should authenticate to each cloud. 
 The simple password authentication
is not a rational approach, since in that case, the cloud provider should share the user password with 
all third-party applications (including \FACADE) granting limitless access to all the applications. 
There are multiple authentication protocols implemented by public cloud providers for access of the third party application. 
For the sake of simplicity, we assume that the cloud storage has deployed Open Authorization (OAuth) authorization framework \cite{oauth} as it is the common approach
that cloud services use to provide client applications with secure delegated access. 
The ``User Authentication'' type defined by OAuth 2.0 \cite{oauth}, uses an ``Access Token'' for a specific user and application pair. Using 
the access token, \FACADE can interact on behalf of the user with the cloud 
to perform the operations (e.g., upload, and download).
\FACADE crucially stores the authentication credentials (e.g., the access token),
specific to each platform on a \textit{password manager}, protected under the user's
single (master) password. Upon requesting access, the user enters the password  to 
get the authentication tokens from the password manager to interact with the storage providers.

\begin{table*}[!t]
	\scriptsize
	\centering
		\vspace{-2mm}

	\caption{A comparison between \FACADE and other related schemes}
	\label{tab:compare}
	\resizebox{1\textwidth}{!}
	{\begin{tabular}{l||l|l|l|l|}
			\cline{2-5}
			& \textbf{Baseline}         & \textbf{Enhanced-baseline} & \textbf{(2, 3)-PPSS}  & \textbf{(2, 3)-PASSAT} \\ \hline\hline
			\multicolumn{1}{|l||}{\textbf{Confidentiality}}       & No                                                                                  & Yes                                                                                   & Yes                                                                                                       & Yes                   \\ \hline
			\multicolumn{1}{|l||}{\textbf{Integrity}}       & No                        & Yes (manipulation of 1 share)                                                                               & Yes (manipulation of up to 2 shares)                                                                                                       & Yes  (manipulation of up to 2 shares)                  \\ \hline
			\multicolumn{1}{|l||}{\textbf{Availability}}       & Yes (availability of 1 out 1)                         &  Yes (availability of 1 out 2)                                                                                   & Yes (availability of 2 out of 3)                                                                                                       & Yes (availability of 2 out of 3)                   \\ \hline
			
			\multicolumn{1}{|l||}{\textbf{Data Deletion Resistance}}       & No                        &  No                                                                                   & Yes                                                                                                      & Yes                    \\ \hline \hline
			
			\multicolumn{1}{|l||}{\textbf{Storage Cost}}       & 1 $\times$ secret size           & 2 $\times$ secret size                                                                       & 3 $\times$ secret size                                                                                           & 3 $\times$ secret size       \\ \hline
			\multicolumn{1}{|l||}{\textbf{Server Transparency}} & Yes                        & Yes                                                                                    & No                                                                                                       & \textbf{Yes}                    \\ \hline                                                                              
			\multicolumn{1}{|l||}{\textbf{Efficiency}}         & Data stored in clear text & Encryption (costly)                                                                           & \begin{tabular}[c]{@{}l@{}}Secret sharing, Encryption, \\ PPSS to reconstruct encryption key\end{tabular} & \textbf{Secret sharing (fast)}  \\ \hline
			\multicolumn{1}{|l||}{\textbf{Token protection}}       & No          & No                                                                       & No                                                                                           & \textbf{Yes}       \\ \hline
			\end{tabular}}
\end{table*}

\noindent \underline{\textbf{\FACADE vs. Other Schemes:}}
To compare our approach with the existing work, we consider as the ``baseline'', a  
cloud storage system that stores the files ($sc$). Such a system does not provide any 
security assurance. However, we can assume an ``enhanced-baseline'' system in which the user duplicates 
the files and store them on 2 platforms for higher availability assurance and also,  
$sc$ could get encrypted on the client-side under an encryption key (which itself 
should be protected by the client). 
For comparison, we also consider the password-protected secret sharing (PPSS) protocol, formalized in 
\cite{bagherzandi2011password}, in which $sc$ is secret-shared and distributed among 
multiple platforms and can be accessed on the basis of a single password $pw$ (by interacting with $k$ 
out of $n$ servers). 
In all cases, the interaction between the client machine (from which the user accesses the files) to 
the cloud storage is assumed to be authenticated with an access token $tk$. 
\FACADE provides the following main properties \textit{simultaneously} 
as presented in Table \ref{tab:compare}: 

\begin{enumerate}[leftmargin=*]

	\item \textbf{Server Transparency:} \FACADE is a client application that interacts with cloud servers 
through the API calls defined by the cloud provider. Therefore, \FACADE can easily get integrated with 
current popular cloud platforms with no modification on the servers. The baseline system is also server transparent, 
as is the case for the enhanced-baseline system since it considers the encryption and replication load to be delegated to the users.
The PPSS system, however, does not provide server-transparency since it requires reconstructing the shares 
through the PPSS protocol running between the client (on input $pw$) and the cloud servers. 

\item \textbf{Efficiency:} Use of the XOR secret sharing of \cite{kurihara2008new} enhances the performance 
of the secret sharing, and therefore, even moderately large files can be secret-shared with minimal 
overhead (as can be observed in our evaluation presented in Section \ref{sec:eval}). Compared to the enhanced-baseline 
system, \FACADE is more efficient due to the lower secret sharing delay compared to the client-side encryption. Also,
in other secret sharing protocols, the delay imposed by the secret sharing algorithm makes it inefficient 
to secret-share large files. Therefore, most of the proposed secret sharing techniques use 
encryption in combination with secret sharing (to distribute the encryption keys) to protect large files 
in the cloud. That is, large files are stored in the cloud encrypted under an encryption key that is 
secret-shared among multiple platforms each having piece of the encryption key.

\item \textbf{Protection of Access Tokens:} \FACADE employs a password
	manager to securely store the tokens on the password manager and
		protect it under the master password to provide secure and easy
		authentication to the cloud service providers.  This approach
		improves the accessibility and security of the tokens.  Storing
		the token on the client application is not secure against
		client compromise.  Besides, it would make it difficult to
		access the cloud storage from different clients because the
		token should be copied on every client from which the user
		wants to access the files.  Note that since the token is
		typically a long string of random characters, it is not
		reasonable to ask the user to memorize it. 
		We can also apply a second layer of secret sharing on the access tokens 
		and share them among multiple password managers to further improve the security 
		by lifting the trust on the password managers. None of the other approaches provide
		this useful feature.

\end{enumerate}

\noindent \underline{\textbf{Our Contributions:}} 
Our novelty lies in building an intrusion-tolerant outsourced
storage system, via a careful combination of known techniques (secret sharing and
password management), that does not require changes to the server side or heavy
encryption on the client side, and is therefore readily deployable.
Our contributions are as follows:

\begin{enumerate}[leftmargin=*]

\item \textbf{System Design: }
We introduce \FACADE, a password-controlled secret file sharing client application, 
that can securely and efficiently store files on public cloud storage services.  \FACADE
employs a light-weight, provably secure XOR secret sharing algorithm to secret-share and reconstruct 
files. Each share of file is stored in clear text on one public storage but does not 
reveal any information about the file, unless a threshold of services get compromised or colludes with one another. \FACADE stores access tokens, required to access/authorize the file operations,
on a password manager protected under the user's master password. To store the file shares and to 
reconstruct the file, the user enters the master password  to unlock the 
tokens stored on the password manager so that \FACADE can 
authenticate to the cloud platforms. 

\item \textbf{\FACADE Implementation as a Client-Side Application: } As its practical instantiation, 
	we develop a \FACADE client application in Python based on (2, 3)-XOR secret sharing.	
		Our application interacts with Google Drive, Box, and Dropbox to store 
file shares and restore them to reconstruct the file. The interaction with the cloud platforms 
is through the API calls defined in the SDKs and the user authentication is based on access tokens
defined by the cloud for the user and \FACADE application. The access tokens are stored in LastPass and \FACADE uses 
LastPass API to fetch them on the master password input from the user.

\item \textbf{Evaluation of \FACADE: }
We evaluated our implementation of \FACADE in terms of average delay incurred to share multiple file sizes, store them on each cloud, 
recollect the shares from the cloud, and reconstruct the file with any combination of 2 out of 3 
shares. Our evaluation shows that \FACADE is efficient and does not impose more that 5\% overhead over the upload and download operation. 
Therefore, it can instantly be incorporated in practice. Based on 
our evaluation, we got insights on the optimum size of the files and share blocks.

\end{enumerate}

\noindent\textbf{Paper Outline:} The rest of the paper is organized as follows. Section \ref{sec:background} provides background on information dispersal and secret sharing.   Sections \ref{sec:system} presents the abstract design of \FACADE. Section \ref{sec:impl} presents a simple proof of concept implementation of \FACADE as a client application. In section \ref{sec:eval} we evaluate the performance of \FACADE file operations. Section \ref{sec:related} scrutinizes the related work. Finally, Sections  \ref{sec:discussion} and \ref{sec:conclusion} discuss the findings and conclude the paper.

	\vspace{-2mm}
\section{Background}
\label{sec:background}

\noindent \textbf{Data Dispersal.}
Several methods are proposed to securely disperse the data  \cite{rabin1989efficient,shamir1979share,krawczyk1993secret,bellare2007robust,bowers2009hail}. 
The secret sharing notion introduced by Shamir and Blakley in 1979 \cite{shamir1979share,blakley1979safeguarding} is a common approach for securely dispersing data. It refers to the method of distributing a secret between a group of participants such that each participant has a portion of the secret and    
only authorized subsets of participants can reconstruct it, i.e., any number of shares fewer than a certain threshold infers no meaningful information about the secret. 

\noindent \textbf{XOR Secret Sharing.} 
Since the introduction of the secret sharing notion several constructions of secret sharing
schemes have been introduced by researchers \cite{benaloh1990generalized,ito1989secret,beimel2011secret}.  However, one major problem with many of the proposed approaches including Shamir \cite{shamir1979share} is the noticeable computational cost for both distribution and recovery \cite{kurihara2008new}. 
In this paper, we build our system based on a relatively new and efficient secret sharing scheme introduced by Kurihara et al. \cite{kurihara2008new}. This  scheme uses light-weight XOR operation for secret sharing and reconstruction. 
 Kurihara et al. \cite{kurihara2008new} (k, n)-threshold scheme enables to make and distribute $n$ shares following the XOR secret sharing algorithm. The secret can be recovered by concatenating $k$ out of $n$ secrets following the reconstruction algorithm.
The XOR operation offers a fast distribution and recovery of secret and reduces the computational cost compared to other schemes with complex operations.
For example, unlike the Shamir's scheme, which uses arithmetic in a finite field F of size P (a prime number) and therefore has a heavy computational cost due to the processing of $(k - 1)$-degree polynomials, the XOR uses light-weight bitwise operations that is much faster.
We used an instantiation of this protocol with $k=2$ and $n=3$, where $n$ is the number of shares and $k$ is the minimum number of shares required to recover the secret.  

\textit{Generating Shares:} At the share generation stage, 
	(2, 3)-XOR threshold scheme divides $d$-bit secret $SC$ into two pieces of $SC_1$ and $SC_2$ each of size $d/2$-bit. For generating the shares we pick a random number $R$ of size $d$-bit.  The $d/2$ higher bits of the first share $c_1$ are created by XORing $d/2$-bit string of zeros with $R_1$ (the higher $d/2$ bits of $R$). The $d/2$ lower bits of $c_1$ are created by XORing $SC_2$ with $R_2$ (the lower $d/2$ bits of $R$). 
The second share ($c_2$) is created by XORing $R_2$ with zeros and $SC_1$ with $R_1$. The third share ($c_3$) is generated by XORing  $SC_1$ with $R_2$,  and $SC_2$ with $R_1$.

\textit{Reconstructing Secret:}
Every two parties can jointly reconstruct $SC$ by performing simple XOR operations.
To recover the secret $SC$ using $c_1$ and $c_3$, a simple XOR of shares results in the production of $SC_2$ on the higher $d/2$ bits and $SC_1$ $\oplus$ $SC_2$ on the lower $d/2$  bits. Another round of XOR between these two ($S_2 \oplus  (S_1 \oplus S_2)$ results in $SC_1$. Swapping the order of recovered $SC_1$ and $SC_2$ blocks result in the reconstruction of the secret $SC$. Similarly the $c_2 \oplus c_3$ produces $S_1 \oplus S_2$ on the lower $d/2$ bits and $SC_1$ on the higher $d/2$ bits. Another round of XORing these two blocks ($(S_1 \oplus S_2) \oplus S_2$) recovers $SC_2$. Finally, a simple $c_1 \oplus c_2$ operation reconstructs the secret $SC$.  

	\section{Overview and Models}
\label{sec:system}



\subsection{System Overview} 
\label{overview}

%

%

\noindent \textbf{Parties:}
Following parties involved in the protocol:

\begin{itemize}[leftmargin=*]

\item $U$: denotes a user who has some information that she wants to store and protect.
\item $S1,...,S_n$: denotes n cloud providers. We let each server $S_i$ pick the access token $tk_i$
\item $C$: is the cloud storage client from which $U$ stores and retrieve $sc$. $C$ interacts with $S$ through the $S$ API calls. 
\item $M$: the password manager that stores access tokens $tk_i$ under the server name $S_i$, and protects them by a single master password $pw$. 

\end{itemize}

User's $sc$ is distributed among $S$'s through a secret-sharing protocol (as is defined next) and can later be accessed by interacting with $k$ out of $n$ servers on the basis of a single password $pw$ and running the reconstruction protocol as follows:

\smallskip
 \noindent \textbf{Secret Sharing:} 
Steps taken to secret-share a file are as follow:

\begin{itemize}[leftmargin=*]

\item Step1-C: Secret-share the secret file $sc$ into shares $(c_1,...,c_n)$
\item Step2-C,M: C interacts with $M$ to retrieve $tk_i$ on input $pw$
\item Step3-C,S: C authenticates to $S_i$ using $tk_i$ 
\item Step4-C,S: C uploads $c_i$ on $S_i$
\end{itemize}

\noindent \textbf{Reconstruction:} 
Steps taken to reconstruct a file:

\begin{itemize}[leftmargin=*]

\item Step1-C,M: C interacts with $M$ to retrieve $tk_i$ on input $pw$
\item Step2-C,S: C authenticates to $S_i$ using $tk_i$ 
\item Step3-C,S: C downloads $k$ out of $n$ shares $c_i$ from each server $S_i$
\item Step1-C: Reconstructs the secret file $sc$ from shares $(c_1,...,c_k)$ 
\end{itemize}

\begin{figure*} [t]
	\vspace{-28mm}
	\centering
	\includegraphics[width=0.9\textwidth]{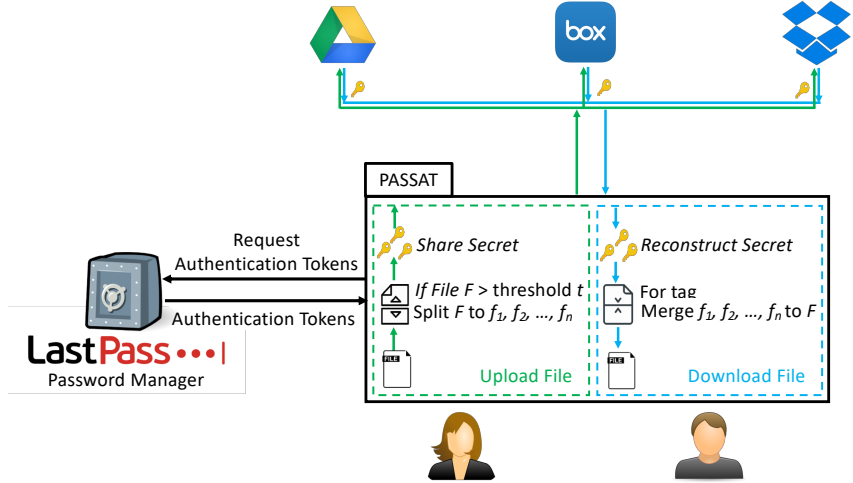}
	\vspace{-30mm}
	\caption{PASSAT high-level view architecture}
	\label{fig:overview}
\end{figure*}

\subsection{Threat Model and Security Properties}
\label{sec:threat}

\noindent \textbf{Threat Model:}
In \FACADE threat model, we consider that each single cloud provider may get compromised through a malicious insider or outsider attack. Therefore, the user should not be obligated to trust any single cloud provider. 
{The data stored on the compromised cloud may be learned, lost, or tampered, temporarily or permanently. 
We assume that an attacker has access to no more than $k-1$ shares, otherwise the attacker can reconstruct the secret file from the learned secret shares, alter the file by modifying/replacing the shares, or
make disruption in accessing the file. Since we do not trust the cloud providers, we assume that no encryption is applied on the files.
We also assume that no more than $k-1$ attackers collude to share the learned secrets. 

We further assume that $M$ is trusted and we consider a secure communication channel between $C$ and $M$ (as is the case when using real-world password manager). 
Note that this property can also be replaced with no trust on the password manager if we add a second layer of secret sharing on $tk$'s to secret-share and store them on multiple password manager. 
We will discuss this solution and other mitigations in Section \ref{sec:discussion}.

\noindent \textbf{Security against Storage Compromise:}
Similar to other (k, n)-secret sharing approaches, we guarantee the security of the system as long as the attacker has access to no more than $k-1$ shares. If each and every $k$ server gets compromised by individual attackers, as long as attackers do not collude, the security is guaranteed. In such a system, the security is compromised only in a situation where an attacker (or more than one attacker who can collude) can retrieve $k$ shares. The system also guarantees that a benign user with access to $k$ shares to be able to reconstruct the whole secret. The availability of the data (and deletion-resistance) is guaranteed if at least a valid set of $k$ servers/secret shares are available. The (k, n)-secret sharing can also protect the integrity of the data (thereby provide modification-resistance) since modified data can be detected by the user while reconstructing the data, as such reconstruction involving modified secret shares will result in a meaningless random output.

\noindent \textbf{Security against Attacks on Password Manager:}
Since we assume that tokens are secured in the password manager  an attacker can only access tokens by launching online guessing attacks against the master password. TThe attacker who has access to the passwords database (being protected under the master password) tries to guess the password to disclose (decrypt) the tokens. After each guessing attack the attacker should then try to access the storage using the disclosed tokens in an online attempt to authenticate to the storage server. Such an attempt may or may not be successful, i.e., the tokens may or may not be valid. However, the mitigation against online attacks is straightforward and has already been deployed by the service providers through rate limiting. Hence, even if the attacker gets access to the password files, obtaining the tokens still requires launching an online guessing attack, which can be detected by the servers. 
Offline attacks on the password file are also not possible since the  stored tokens have high entropy and cannot be guessed, unlike a password driven from a known dictionary. 


%

\section{System Design} \label{sec:overview}

 \noindent \textbf{\FACADE Components. }
\FACADE consists of three main and one optional components:

\begin{enumerate}[leftmargin=*]

\item \textit{Secret Generator.} This component receives a secret file as an input and secret-shares it into $n$ shares. 
It also reconstructs the file from at least $k$ out of $n$ shares. 

\item \textit{Middleware.} This component act as the interface between the user, the password manager, and the cloud services. 

\item \textit{Password Manager.} The password manager is used as a secure authentication token vault that stores access tokens needed to authenticate \FACADE to the clouds. \FACADE authenticates to the password manager using the user (master) password. 

\item \textit{Splitter.} XOR operation can be daunting and can slow down the process for large files. Therefore, \FACADE optionally deploys a file splitter that can break a large file into smaller files and tag them for the later reconstruction of the file. 

\end{enumerate}

%
%

Figure \ref{fig:overview} demonstrates the overview of \FACADE. We assume that the user has accounts with $n$ public cloud services. Optionally \FACADE can be provided as a service with no need for the user to register an account with each storage service. 
Each uploading file is secret-shared and each share is stored on one of the $n$ cloud services. \FACADE can split larger files into smaller files before secret-sharing (details is discussed in Section \ref{sec:impl} and Section \ref{sec:eval}). The download procedure starts by receiving the shares from the storage and reconstructing the file following the secure reconstruction protocol. 
The long-lasting authentication information is stored in the password manager once and can be retrieved any time the user needs to upload/download files. To fetch the authentication information from the password manager the user should enter the (master) password, which unlike the long-lasting authentication information could be updated frequently. We can assume that an authentication session can remain active for a limited duration of time (e.g., as long as the \FACADE application is open). 

Figure \ref{fig:send} demonstrates the  \FACADE operational diagram for uploads.  The process starts by the user entering a file name to be stored securely on the cloud. \FACADE receives the file and using the Secret Sharing component  distributes it into $n$ shares. Middleware authenticates the system to the cloud services using the fetched credentials and uploads the shares into the cloud services. If the size of the file is larger than a certain limit, \FACADE can optionally split it into smaller files to improve efficiency. 

Figure \ref{fig:ritrive} demonstrates the operational diagram for downloads. When the user requests a file to be downloaded, \FACADE Middleware authenticates to $k$ out of the $n$ services and downloads the previously stored shares. The Middleware passes the received shares to the Secret Sharing component to reconstruct the file.

\noindent \textbf{Secret Sharing. } \label{sec:sharing}
This component is responsible for generating shares from the secret file and reconstructing the secret file from the shares. The component employs (2, 3)-XOR threshold scheme (discussed in Section \ref{sec:background}) to construct and distribute three file shares out of a file in such a way that at least two file shares are required to reconstruct the original file and having access to one of the file-shares does not reveal any information about the original file. 
For the process of generating shares, the component reads through the file and chops the file into blocks of $d$-bits, applies the (2, 3)-XOR secret sharing on each block, and writes the outputs into three different shares/files. In cases where the file size is not a factor of $d$-bits the component pads the file tail with a string of zeros. 
For the reconstruction, \FACADE takes two shared files as the input and reconstructs them in blocks of $d$-bit. The possible padding is removed after reconstruction.

\begin{figure}[t]
	\centering
		\centering
		\includegraphics[width=0.5\textwidth]{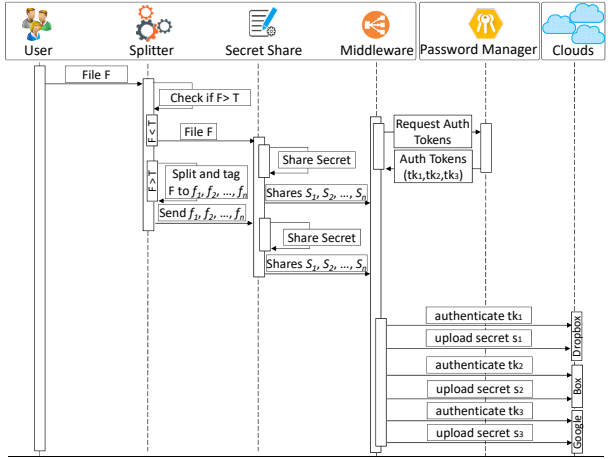}
		\vspace{-6mm}
		\caption{\FACADE components interaction on uploading a file}
		\label{fig:send}
\end{figure}

\begin{figure}[t]

		\centering
		\includegraphics[width=0.5\textwidth]{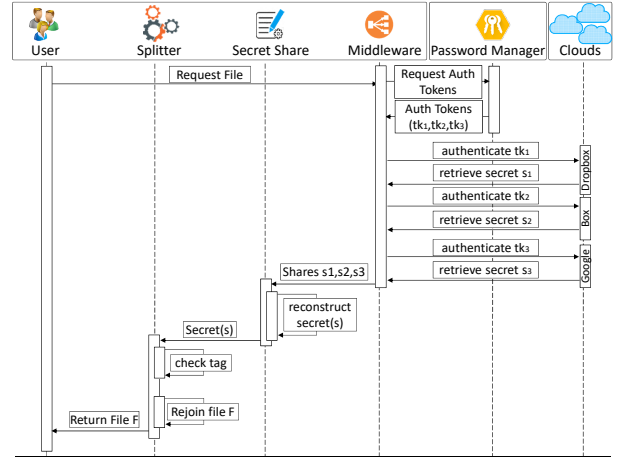}
		\vspace{-6mm}
		\caption{\FACADE components interaction on retrieving a file}
		\label{fig:ritrive}

\end{figure}

%
%

 \noindent \textbf{Middleware. } \label{sec:middalware} 
As a central component Middleware interacts with cloud storages and the password manager, and offers a user-interface, which allows users to upload and retrieve their data. This component uses the APIs and received master password from the users, and retrieves the cloud tokens for further interaction with storages.

%

 \noindent \textbf{Single Authentication Through a Credential Vault. } \label{sec:singlauth}
To list, upload, or download files on popular commercial cloud storage, often the cloud provider requires the apps to authenticate using a long-lasting authentication token (as 
will be elaborated in Section \ref{sec:impl}). \FACADE  just like any other app that uses the cloud API to perform file operations requires to authenticate itself to get access 
to the file directory. One option is for the user to input the credential information at every access. Since these tokens are often a series of long and randomized characters, it would not be 
feasible (or usable) to ask the user to provide them for each cloud storage per each access request. Another option is to store the long-lasting tokens on \FACADE (e.g., hard-
coding the tokens on the app), however, in that case any unauthorized access (even temporarily) to the \FACADE app may lead to exposure of the tokens and thereby access to the 
files from the same \FACADE client or any other maliciously developed app. 
Also since the user may access the files from multiple clients, the access tokens should be copied on each client and synchronized if updated, which increases the security risk. 
Hence, we store the authentication information in the password manager.
%



\noindent \textbf{File Splitting. } \label{sec:split}
Over the past years, the secret sharing schemes have  been mostly used in the context of small secrets, mainly to distribute secret keys and to protect them securely. Computational cost of sharing and reconstructing secrets makes it infeasible to use this technique on large files. To overcome this problem, first, we used a fast and light-weight secret sharing scheme \cite{kurihara2008new}. The scheme utilizes a simple and low cost operation of XOR for constructing and reconstructing shares and secret. Compared to the traditional approaches (e.g., Shamir's secret sharing schemes \cite{shamir1979share}) XOR requires a significantly less computational time and poses considerably less overhead. Second, for the further enhancement and to avoid impelling huge computation of XORing large file, we designed a splitter that breaks larger files into smaller files. Breaking larger files into smaller files accelerates the process whenever the file size passes a specific threshold, which may impact the secret sharing computation time (as will be discussed in Section \ref{sec:eval}). Splitting a large file into small files enables the system to reduce the computational time by dealing with smaller files. For future reconstruction Splitter tags each of the new files. Assuming that we have a finite file \textit{f}, the splitter split the file into $\{f_1,f_2,f_3,...,f_t\}$. These smaller files later are the input to the secret sharing component. Figure \ref{fig:files} shows an overview of breaking a file.

\begin{figure} [!t]
		\vspace{-20mm}
	\centering
	\hspace*{1mm}	\includegraphics[width=1\textwidth]{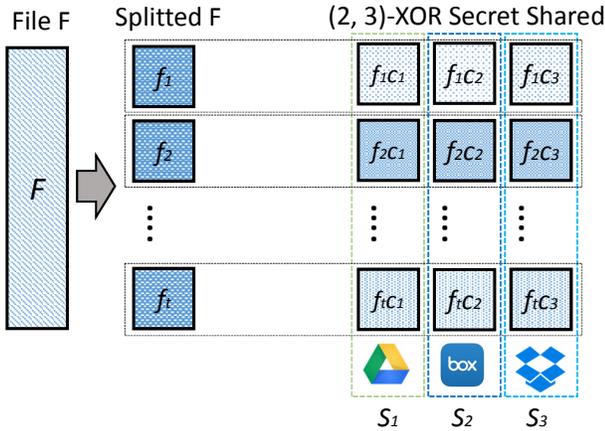}
		\vspace{-66mm}
	\caption{Splitting a file to smaller files and sharing them across different cloud services}

	\label{fig:files}
\end{figure}

\section{Implementation}
\label{sec:impl}


Although \FACADE can be deployed with any other number for $k$ and $n$, we 
prove the practicality of our approach with $k = 2$, and $n = 3$ (tolerating compromise or unavailability of one service at a time). 
Appendix Algorithms \ref{secretshare2} and \ref{reconstruct2}  delineate \FACADE approach on sharing and reconstructing secret. 
Considering the current server level agreement and cloud uptime the (2, 3) seems to be a resilient threshold parameter and offers an acceptable level of availability. These parameters
have also been adopted before (e.g., \cite{veeam,123backup,titan,backblaze}). 
While the security enhances with the increase in $k$ and $n$ (in terms of making the system more reliable against possible cloud unavailability and/or compromise), 
the storage cost and shares computation time grows as $k$ and $n$ increase, which may make the system infeasible for larger files in practice.

We developed \FACADE as a client-side application in Python. We used the libraries offered by the cloud storage services and password managers in our deployment. 
In our implementation, during the setup phase, the user requires to register with the desired cloud services, namely, Google Drive \cite{googleapi}, Box \cite{boxapi}, and Dropbox \cite{dropboxapi}, and obtain the access tokens required for further communication with the cloud only. In this proof of concept deployment registering the users was a 
simple implementation choice, however, storage providers offer several other authentication mechanisms for third party apps that do not require user registration to support different use cases. 
Alternatively, we can assume \FACADE as a service provider which generates and manages the accounts and interacts with third party storage service providers as will be discussed in Section \ref{sec:discussion}. 
\FACADE does not record the user's credentials locally by any means. Instead, the user stores the access tokens on LastPass password manager \cite{LastPass} under the server's name during the initial setup. To securely communicate with the storage, \FACADE fetches the access tokens from the password manager using LastPass Python libraries. The access tokens are used by \FACADE to securely connect to the cloud and perform file operations (e.g., listing, uploading and downloading files from the cloud). 
As mentioned earlier, for the sake of simplicity, we assume that the password manager is trusted and store the tokens only on one manager.

After setting up the system (registering with the storage services and storing the tokens on LastPass), the user can perform the file operations by inputting the file operation type (upload, or download), the file address for uploading, the file name for downloading, and the (master) password to connect to LastPass to \FACADE.  
For the file upload operation, \FACADE works as follows, first, the program decides whether it is required to break down the larger file into smaller files. The smaller files are tagged by adding a sequential number to their name for easier merge during the download procedure. The decision is made based on a dynamic threshold which can be defined to reduce the computational time for generating and reconstructing the secret. Our evaluation section gives some insights on the optimum file sizes. \FACADE, continues with secret-sharing the file by processing the file block by block and applying the (2, 3)-XOR secret sharing discussed in Section \ref{sec:background} on each block of size $d$-bit. 
Note that due to the underlying hardware and processing specification, running the secret-sharing on a file of size $d$-bit is not efficient. Therefore, as will be discussed in our evaluation the size of $d$ is picked based on the optimum performance. 
\FACADE appends the $d$-bit shares to form each file share to be stored on the cloud services. 
The system then uses the provided LastPass password to fetch the access token for each of the three storage services and then makes the API calls to each cloud using the access token to store the file. 
Similarly, for the file download operation, \FACADE communicates with any 2 of 3 cloud storage services using the authentication token received from LastPass to get the file shares. \FACADE applies the (2, 3)-XOR reconstruction procedure on the received file shares block by block and appends the result to reconstruct the file. 
For larger files, \FACADE receives all the split file shares, reconstruct the split files by applying the (2, 3)-XOR reconstruction procedure, and then merges all the reconstructed shares to form the actual file. 
Following sections describe this procedure in detail.

\noindent \textbf{Secret Sharing.} \label{secretsharing}
This includes two main functions of construction and reconstruction written in Python. The construction function reads the file in $d$-bit block and writes them into the shares. The $d$-bit block is appended to form a file share that is stored in the cloud. Reconstruction rebuilds the file from any two shares. The file shares are processed at $d$-bit block sizes and outputs are appended to reconstruct the file. 

%
%

\noindent \textbf{Interaction with Cloud Services.} \label{cloudsapi} 
\FACADE generates three files shares, each one should be stored in one of the three cloud storages, namely,
 Google Drive, Box, and Dropbox. 
We store each of these file shares using the APIs provided by each of these companies. 
We used the Drive API V3 through \texttt{google-api-python-client} Python package version 1.6.4  for interaction with Google Drive. 
Our Google Drive upload function uses \texttt{MediaFileUpload} to define the file and 
runs an HTTP POST request to \texttt{create} the file on Drive. Our Google 
Drive download function creates an HTTP GET request to download the file.  
We used Box API through \texttt{boxsdk} Python package version 2.0.0a11  for operations on Box platform. 
Communication with Box API is similar to Google Drive through HTTP POST and GET requests. 
For Dropbox, we used the Dropbox API V2 through \texttt{dropbox} Python package version 8.5.0  to upload and download files. 
Similar to the Google Drive and Box API we make HTTP POST and GET requests to upload and download the files.

%
%
%
%
%
%

\noindent \textbf{LastPass.} 
We used LastPass \cite{LastPass} as a popular password manager and based the development
 on \texttt{lastpass-python} package to interact with LastPass and leverage the provided functionality. 
Our current implementation setup requires users to login and store their access tokens into their local LastPass application,
though the process can be automated as an additional option. 
Once the credentials are stored, \FACADE uses Lastpass API to retrieve the cloud access tokens.

\noindent \textbf{Splitting and Merging.} 
We deployed a simple and fast file splitting to break larger files $f$ into smaller files using a threshold that can be defined based on the performance of the operations as will be discussed in Section \ref{sec:eval}. We created a Python function, which reads through the file and splits and tags it for the later reconstruction. While our current splitting function is only able to manage the text files, as a part of future enhancement we aim to expand and empower this function to handle other file types.  

	\begin{figure} [!t]
	\vspace{-8mm}
	\centering
	\includegraphics[width=0.45\textwidth]{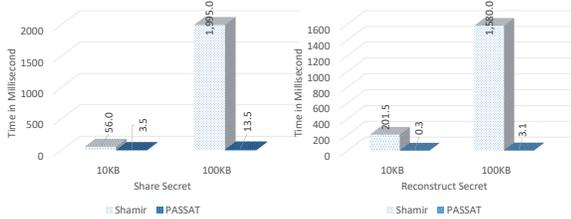}
	\vspace{-26mm}
	\caption{Shamir secret sharing and PASSAT Latency comparison}
		\vspace{-6mm}
	\label{fig:shamus}
\end{figure}

\section{Evaluation} 
\label{sec:eval}

Prior studies have discussed and evaluated the performance of different schemes \cite{shor2018best,junghanns2016engineering,kurihara2008new,resch}.
\cite{kurihara2008new} compared the XOR  and Shamir's secret sharing. \cite{resch} compared their system with Rabin and Shamir approach.  \cite{shor2018best} and \cite{junghanns2016engineering} reported detail comparisons of various approaches. 
We further compared our approach against the popular Shamir's scheme which seems to be the fastest schemes \cite{junghanns2016engineering,shor2018best}. Figure \ref{fig:shamus} shows the differences in terms of latency between two approaches. We picked two file size of 10KB and 100 KB as any computation (1MB and 100MB) result in a significant increase in Shamir's secret computation. As it can be seen our approach has far less latency compared to the popular Shamir's approach.
We briefly mentioned  the heavy computational cost of Shamir's secret sharing in contrast with XOR (Sec.~\ref{sec:background}) and 
provided a detailed qualitative evaluation between baseline, enhanced baseline, and \FACADE (Table \ref{tab:compare}). Here, we analyze
the performance of \FACADE in terms of the \textit{``storage cost''} and \textit{``latency''}. We evaluate the overhead compared to the baseline (uploading and downloading files) to prove the viability of such approach and demonstrate that the security benefits come at a reasonable cost. We have not performed any code optimization, therefore, the delays reported in this section are the upper side estimation of the delay. Since we do not have any server-side changes we consider the storage providers as a black box and do not have any evaluation of the internal processes of the cloud services. 
Therefore, for the baseline system, we only measured the interaction time between our application and the cloud storage to demonstrate the average delay of file uploading and downloading. 
We ran our test on a 64-bits Windows platform (6.1, Build 7601) running on a 8.00 GB (7.87 Usable) RAM, and an Intel(R) core(TM) 2quad
q9550 @2.83 GHz processor representing a client machine. 
We evaluate and report the delay of different operations averaged over 1000 iterations. 
During our test, we used a 1 Gbps network link and monitored the stability of the  connection to avoid any possible interruption or fluctuations.  

\noindent \textbf {\FACADE Storage Cost and Latency.} 
The storage required to store a data in the baseline situation with no security and availability guarantee is equal to the size of the file.  
 However, most of the common approaches recommend having at least 3 instances of a data for backup procedure \cite{123backup}. 
 If we assume having three backups of a data as an enhanced-baseline that offers availability (similar baseline were assumed in other studies, e.g.,\cite{wylie2000survivable}) \FACADE requires exactly the same amount of storage to store data securely while guaranteeing the confidentiality. 
For instance, \cite{wylie2000survivable} requires 6 times of the file size storage  to guarantee the confidentiality and availability. The scheme proposed in \cite{bessani2013depsky}, requires four times of the file size storage in DEPSKY-A and over 2 times in DEPSKY-CA. PPSS \cite{jarecki2014round} with the (2, 3)-secret sharing same as \FACADE utilizes three times of the file size storage. 


\noindent \textbf{Baseline Latency. } 
To measure the latency, we replicated the process of uploading and downloading files to and from each of the cloud providers individually and considered that as the baseline. 
This time does not include the authentication time to the cloud storage service. 
Therefore, we consider our baseline as the most optimistic scenario which assumes that the user is already authenticated to the services. 
Figures \ref{fig:upload} and \ref{fig:download} represent the average time it takes to upload and download different file sizes into and from each of these cloud providers.
As for the upload, as expected the minimum delay is related to the small files of size 10KB with an average of about 1.6 seconds. For larger files of size 200MB, the delay is about 40 seconds. The download takes about 0.7 seconds for the small size of 10KB and about 55 seconds for the large file size of 200MB. We do not intend to compare different platforms with each other, however, the difference between them is mainly related to the API implementation. For example, Google Drive uses a \texttt{resumable} download for files larger than 5MB and downloads the files chunk by chunk, which is perhaps reflected in the higher download time from Google Drive on large files compared to Dropbox and Box.

%
%
%

\begin{figure}[!t]
	\vspace{-10mm}
	\centering
	\begin{minipage}{0.48\textwidth}
		\centering
			\includegraphics[width=1\textwidth]{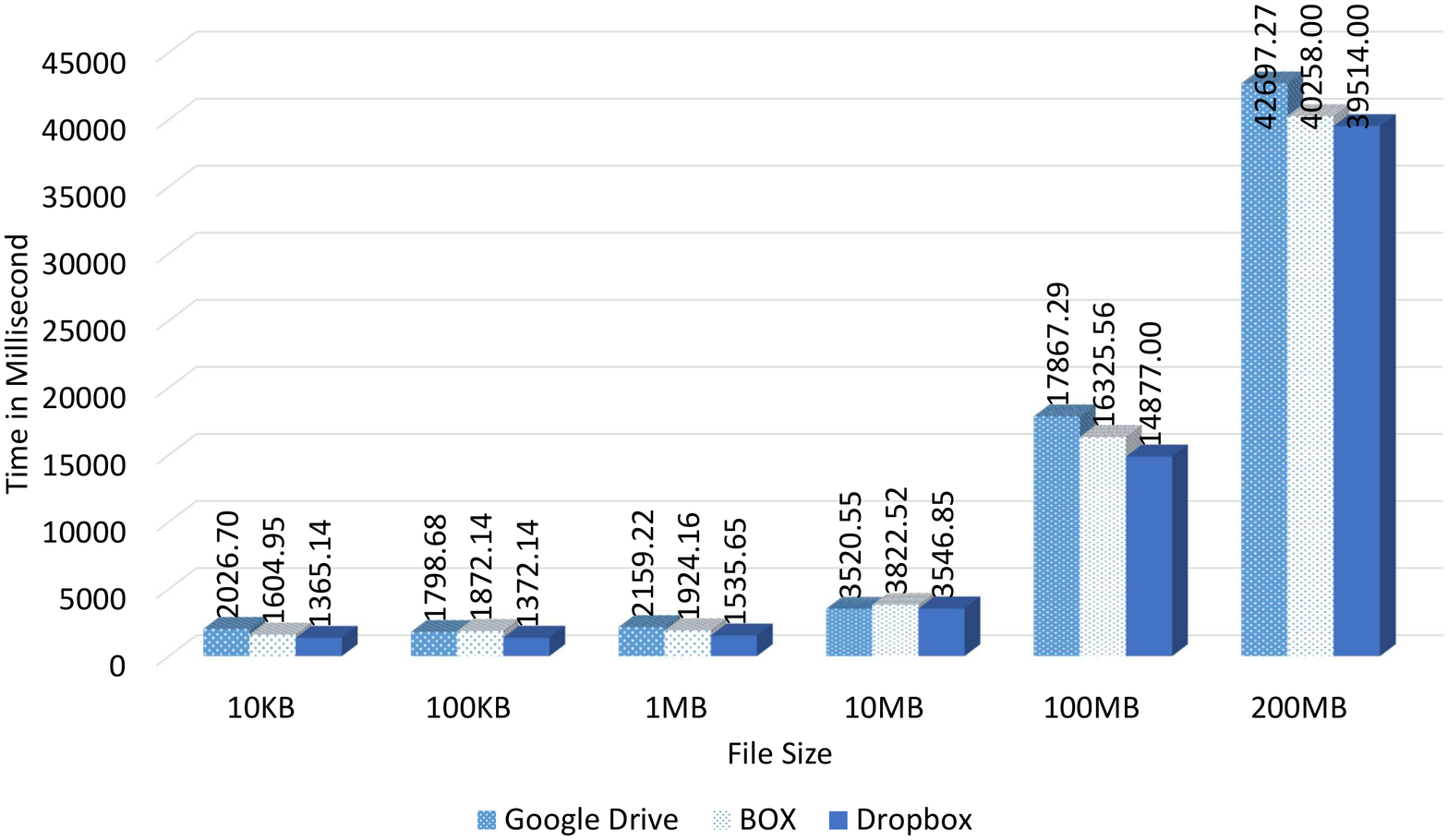}
		\vspace{-12mm}
		\caption{Time to store various files  in major cloud providers}
		\label{fig:upload}
	\end{minipage}\hfill
	\begin{minipage}{0.48\textwidth}
		\centering
	\includegraphics[width=1\textwidth]{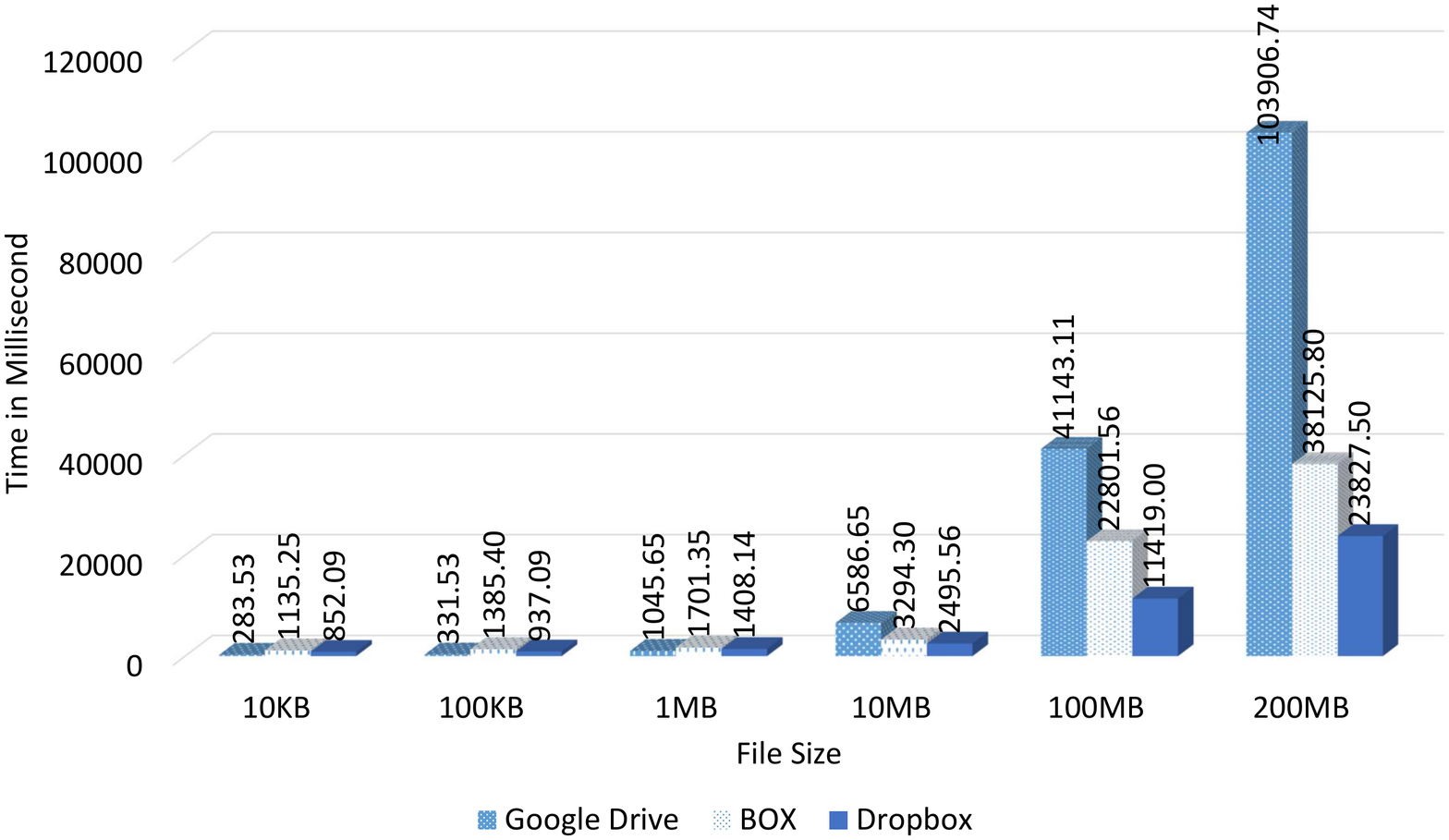}
		\vspace{-12mm}
	\caption{Time to retrieve files from major cloud providers}
	\label{fig:download}
	\end{minipage}

\end{figure}

%

%

\noindent \textbf{Block Secret-Sharing Latency. } 
We measured the time it takes to secret-share and reconstruct each block of size $d$-bit for 5 different block sizes (i.e., $d=2^t, 8 \le t \le 13$). 
We computed the time it takes to secret-share files as the overhead of \FACADE over the baseline. 
We considered small files of size 10KB and 100KB, medium size files of size 1MB, and large files of size 10MB. For each file, we considered 5 different block sizes 
(i.e., $d=2^t, 8 \le t \le 13$).
We computed the execution time averaged over 1000 iterations for each operation. As shown in Figure \ref{fig:blocksizetiming}, overall the (2, 3)-XOR secret sharing 
has a very low latency (few nanoseconds) for various block sizes.

\begin{figure}[!t]
			\vspace{-10mm}
	\centering
	\begin{minipage}{0.48\textwidth}
		
		\centering
		\includegraphics[width=1\textwidth]{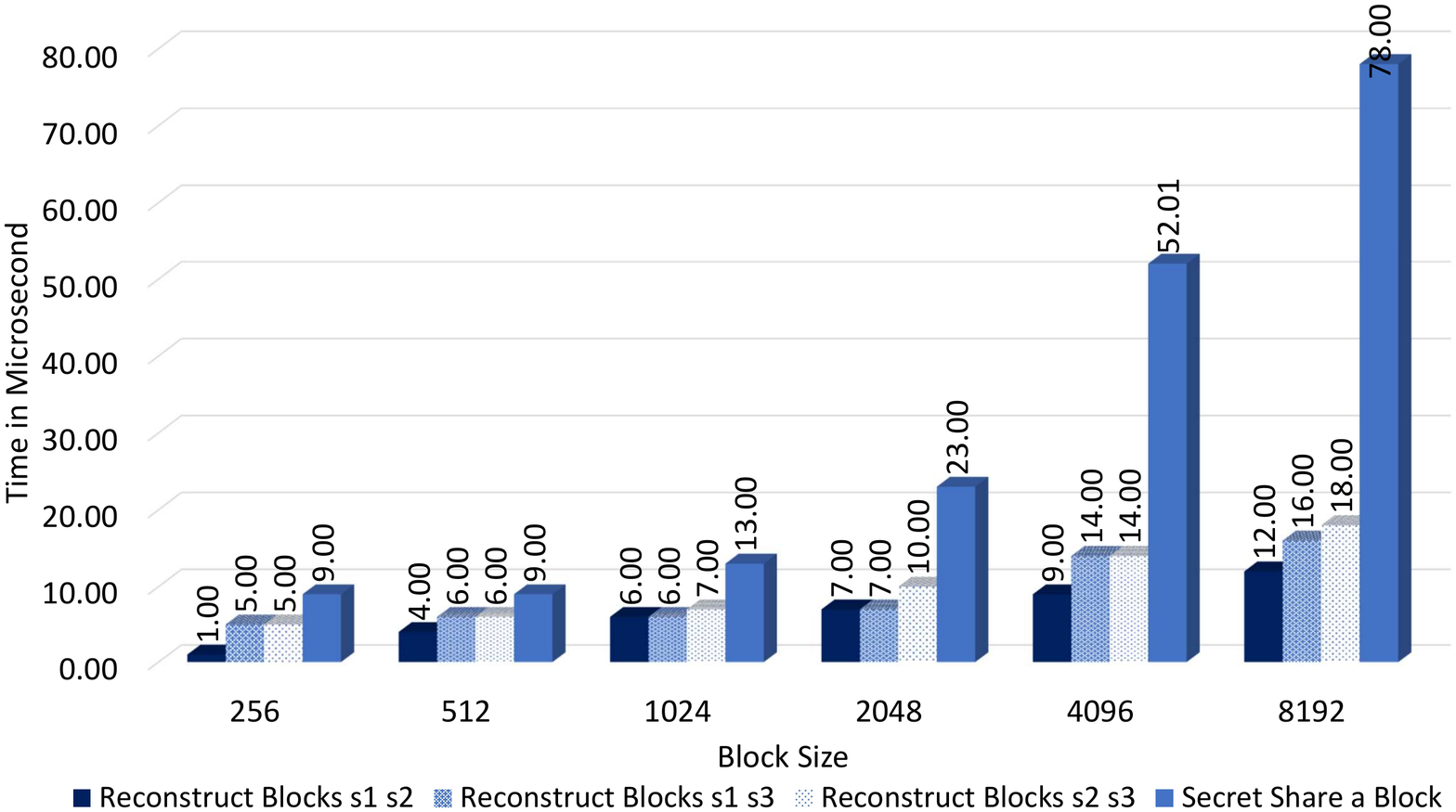}
		\vspace{-12mm}
		\caption{Time to secret-share each block for various sizes}
		\label{fig:blocksizetiming}
	\end{minipage}\hfill
	\begin{minipage}{0.48\textwidth}
		\centering
		\includegraphics[width=1\textwidth]{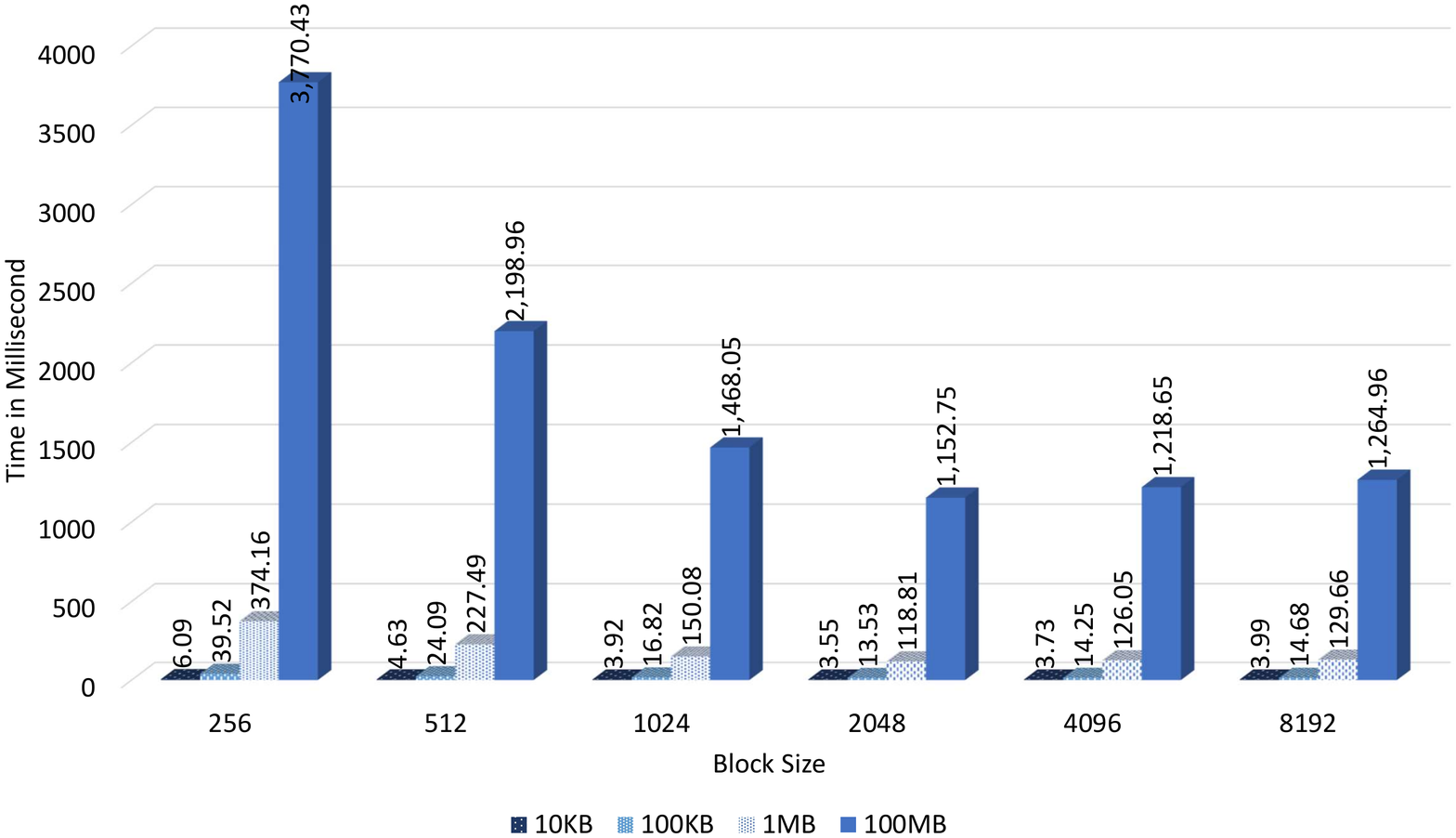}
		\vspace{-12mm}
		\caption{Time  to secret-share various files using various sizes }
		\label{fig:secretsharetiming}
	\end{minipage}
\end{figure}

\begin{figure*} [!t]
	\vspace{-18mm}
	\centering
	\includegraphics[width=1\textwidth]{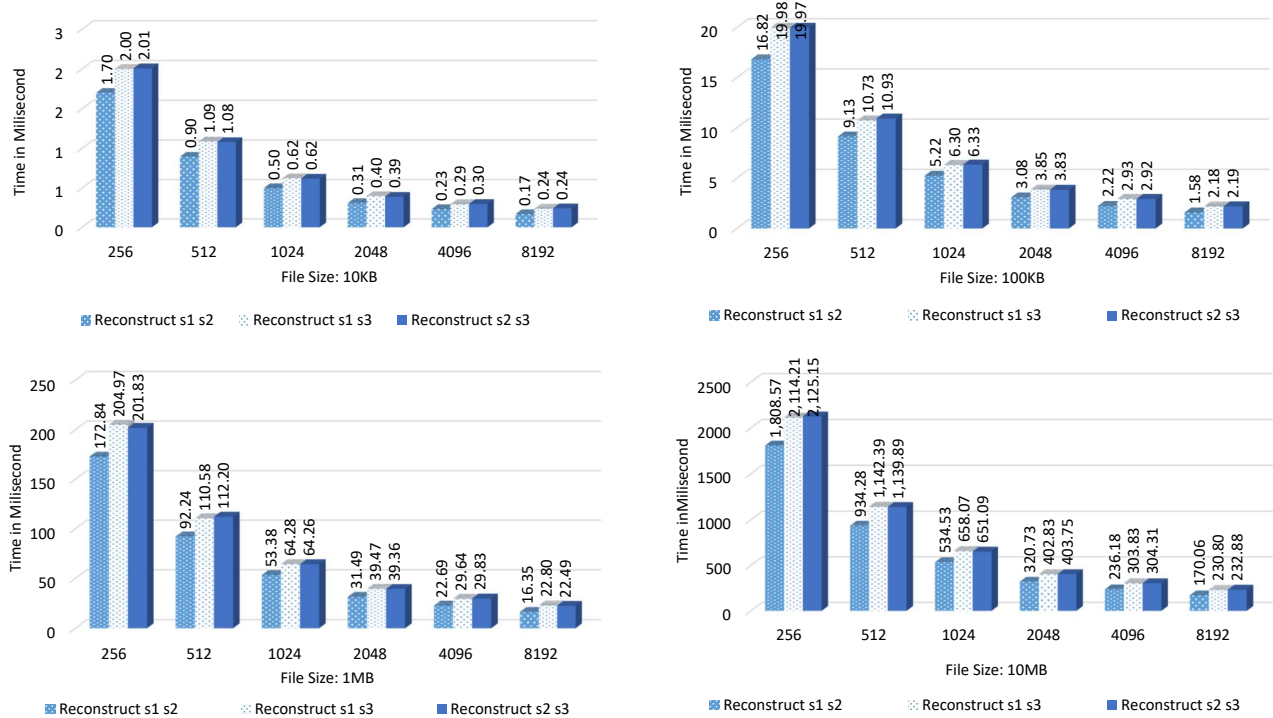}
	\vspace{-18mm}
	\caption{Time taken to reconstruct various file sizes with various block sizes from different shares}
	
	\label{fig:reconstructall}
\end{figure*}

\noindent \textbf{File Secret-Sharing Latency. } 
As discussed in Sections \ref{sec:sharing},  \FACADE reads through the file in blocks of $d$-bits and applies the (2, 3)-XOR secret sharing scheme to each block and then appends them to secret share the file. The process time varies depending on the file size and the block size. To obtain the most efficient time for generating the shares and reconstructing the file, we tested \FACADE against different file sizes and using various $d$-bits blocks sizes, aiming to obtain the most efficient block size.
In this computation, we have not optimized the code (i.e., the file is read sequentially $d$-bit at a time and computation happens in one thread).

Figure \ref{fig:secretsharetiming} demonstrates the time taken to generate the shares, which shows an increase as the files size increases. 
If we allow the secret sharing to add a 10\% overhead over the baseline system, we notice that in the file sizes of $\leq$ 1MB, the overhead is 
acceptable ($<$ 5\%). This observation does not stand for the 10MB file. This examination gives us an insight on the optimum size of the file for secret sharing before \FACADE decides to split the file.
We also noticed that regardless of the file size the block size of $d=2048$ offers the fastest secret-sharing time. It seems for block sizes that are larger than $d=2048$ (i.e., in our test 4096, and 8192 bits) the time of secret sharing increases. We relate this to the underlying client processor and memory architecture.

Figure \ref{fig:reconstructall} shows the average time it takes to reconstruct the file. Note that usually a file is downloaded more frequently than it is uploaded (or updated). Also, the network downstream bandwidth is typically higher than the upstream bandwidth. Therefore, the time it takes to recover the file is as of more importance compared to the time it takes to generate the shares from the file.  
Figure \ref{fig:reconstructall} shows that, as the size of the block increases the time it takes to reconstruct the file decreases, the best reconstruction performance happens for the blocks of size $d=8192$ (slightly lower than $d=4096$). 
The reconstruction overhead seems to be acceptable for files of various sizes since the reconstruction only deals with computation on 2 out of 3 shares while during the secret sharing three shares should be generated. Of course, the code optimization and parallelization can improve the delay associated with each operation.

 \noindent \textbf{File Splitting Latency. } 
As mentioned earlier \FACADE can split large files before secret sharing (and merge them back after reconstruction). We developed a simple file splitting of textual file types and computed the average time to split and merge the files. Using this code, splitting a 100MB file to 100 smaller files of size 1MB took on average about 128 ms, and merging the 10 files of size 10MB to the 100MB file took 112 ms. This result shows that splitting does not impose a significant delay and file splitting before secret sharing can be feasible to secret-share large files, given that multiple small files can be secret-shared in parallel (as the operation on small files are independent).

	\section{Discussion and Future Work}
\label{sec:discussion}

\noindent \textbf{Other secret sharing approaches.  } 
In this paper, we built our system using three major free storage providers, however, this approach can be integrated with other cloud providers. 
We also chose a secret sharing scheme which uses simple XOR operation to construct and reconstruct shares. 
The modular design of \FACADE allows the use of other secret-sharing algorithms (e.g.,\cite{shamir1979share,rabin1989efficient,krawczyk1993secret}). Since the secret-sharing component is independent of other modules (e.g., cloud storage and password manager) replacing it with another scheme does not affect the protocol. Similarly, our fundamental idea can work with any manager or storage.

\noindent \textbf{Vendor lock-in. } 
Several works has discussed the cost of data migration between cloud services and the Vendor lock-in problem \cite{abu2010racs,bessani2013depsky,bowers2009hail}. \FACADE addresses this concern since it does not depend on a single provider. Also, any possible rate inflation by any of the providers the data can be recovered and accessed from the two other storage providers. 
Hence, the migration cost can be completely eliminated by leaving the costly storages. That is, in many of the cases \FACADE offers free of cost migration, 
and in cases where two of the vendors increase their
price, only one migration is required. 

%
%
%
%
%



\noindent \textbf{\FACADE as an online service. }
Our current implementation of \FACADE requires users to register with cloud providers and share their credentials with the password manager. However, an efficient scenario is to implement \FACADE as a service to offer reliable storage functionality to users. Therefore, as a part of future work, we aim to deploy \FACADE as a service and study any possible challenges (e.g., security of authentication to the \FACADE service, communication with storages, cost of deployment). 
Further research could also usefully expand \FACADE as a mobile application allowing users to securely share their file from their smartphone. This application can also uses \FACADE web-service to perform file operation. 

 \noindent \textbf{Trust on password manager. }
We discussed how the access tokens could be stored on the password manager, which requires trust on the password manager. To reduce the trust on this component, we can secret-share the access tokens on multiple password managers. Other type of secure storeless password managers such as the one proposed in \cite{shirvanian2017sphinx} could also be used. This type of password manager does not store the password, but reconstructs it through an oblivious cryptographic protocol. Therefore, if \FACADE uses this type of password manager, the tokens can be reconstructed on every session, without the need to store them. However, this approach is valid only if the storage provider allows the tokens to be generated by external entities (not by themselves, as is the case currently with most providers). 
We should also mention that inputting the (master) password is not necessary for each and every operation. The password can be entered for one session and be valid for a period of time.  Similarly, the tokens can be stored by \FACADE for 
without interacting with the password manager. 

	\vspace{-4mm}
\section{Related Work}
\label{sec:related}

Many of the previous researchers focused on availability of the data on the storage. Some considered the Redundant Array of Inexpensive Disks (RAID) as a data storage virtualization technology to combine multiple hard disk components to guarantee data redundancy \cite{patterson1988case,patterson1989introduction,gibson1992redundant}. Depending on the required level of redundancy, they scatter the data over the drives. However, RAID only provides sector redundancy dynamically across hard-drives and protects against data loss by mirroring the driver to get a secure redundancy, and it still does not offer a solution to rectify security and privacy concerns. 
Other studies used the secure erasure code  \cite{abu2010racs,lin2010secure,bowers2009hail,stefanov2012iris,buasescu2012robust}
to enhance the availability of the data.  These methods only target the availability (not confidentiality). Thus, an unauthorized access to any disk can lead to data leakage. 
  
A variety of techniques are introduced to disperse data and improve the availability, security, and privacy of data over the distributed storages \cite{bowers2009hail,wilcox2008tahoe,lin2010secure,popa2011enabling,shraer2010venus,abu2010racs,stefanov2012iris,bessani2013depsky,kamara2011cs2,junghanns2016engineering, wylie2000survivable}.
Approaches like \cite{popa2011enabling} assume that the user encrypts data prior to uploading to the cloud and therefore leave the confidentiality up to the user to address. 
Many of these studies including \cite{lin2010secure,bessani2013depsky,spillner2011information} use a form of encryption to ensure the confidentiality of the data. Apart from the heavy computational cost of encryption and decryption, they use secret sharing to share the encryption keys, which adds another level of complication.
Many of the proposed approaches including \cite{kamara2011cs2,popa2011enabling,bowers2009hail,hendricks2007low,jarecki2014round} require some sort of modification or code execution at the server side, while \FACADE is transparent to the server and requires no additional modification and changes in cloud logic.
Similarly, \cite{lin2010secure} aimed to enhance the privacy of the stored data in the distributed storage systems. They combined a decentralized erasure code and threshold public key encryption scheme. Their cloud storage system is not secure against a probabilistic polynomial time attacker with a non-negligible advantage. Moreover, such a system adds a significant overhead compared to our approach.

HAIL \cite{bowers2009hail} is another approach that ensures the availability and integrity of data through distributed cryptographic systems. HAIL utilizes Proofs of Retrievability (PORs) and relies on a single trusted verifier to verify the integrity of a retrieved file. The system does not use any secret key for the verifiability.  
HAIL offers the integrity and availability by aggregating cryptographic protocols and erasure codes. However, HAIL requires server-side changes.
 More importantly, it does not guarantee the confidentiality of the stored data.
Several studies have been conducted to offer confidentiality, availability, and integrity by combining different techniques \cite{bessani2013depsky, spillner2011information}. Bessani et al. \cite{bessani2013depsky} proposed DepSky, which uses encryption and secret sharing techniques to store the data in different cloud services. DepSky encrypts the data using a random secret key.  It divides the secret key and then distributes block of the encrypted data and a share of the key in each cloud server. However, what makes our study different from \cite{bessani2013depsky} is that they followed the traditional approach of encryption which requires a secure server to manage the keys. Instead, \FACADE requires no server side changes and utilizes a password manager for authenticating to the cloud.   Moreover, DepSky uses the approach proposed in \cite{krawczyk1993secret}, which makes the data only computationally secret rather than theoretically being secure. 
Unlike \cite{bessani2013depsky}, we do not use any form of encryption. Apart from the extra workload that it may cause, our approach is obliging if encryption is prohibited  \cite{CryptoLaw}. 

Also, relevant to our work is the idea of password protected secret sharing (PPSS) generalized in \cite{bagherzandi2011password}. PPSS is a  protocol suggested to address the security and reliability  
issues in storing a secret data on untrusted platforms. In this protocol, a user Alice $U$ has some secret information $sc$ that she wants to store and protect, and be able to later access on the basis of a single password $pw$ from her client machine $C$.  
Note that $pw$ is password specific to the PPSS protocol. The system still requires an access token $tk$ (stored on the client) to authenticate a pair of $U$ and $C$.
In PPSS, $C$ uses $tk$'s to communicate with every cloud server $S_1,...,S_n$ after which, each server $S_i$ stores some information $w_i$ associated with U ($w_i$ is a function of the secret $sc$, the password $pw$ and server name $S_i$). On the reconstruction phase, when $U$ needs to retrieve $sc$ through $C$, she performs a reconstruction protocol by interacting with a subset of at least $k$ servers where the only input from $U$ is her password $pw$ and $tk$'s. 
An attacker breaking into $k-1$ servers (e.g., either a malicious insider or an attacker who knows $tk$) cannot gain any information on $sc$ other than by correctly guessing $pw$ and running an on-line attack with it.
%
%
%
%
%
While PPSS \cite{bagherzandi2011password} can address the availability and secrecy concerns,
unlike our approach, PPSS critically requires server-side changes and thus cannot be used readily with any of the existing storage service providers (thus presenting a major deploymental hurdle). 
Same security properties apply to \FACADE. However, \FACADE (with standard password manager, like Lastpass) does not require server side changes, while in PPSS the server takes part in running the PPSS protocol. 
Also, \FACADE provides a higher efficiency since the XOR secret sharing used in \FACADE is much faster than the encryption, secret sharing, and PPSS protocol. 

	\vspace{-4mm}
\section{Conclusion}
\label{sec:conclusion}

In the paper, we presented \FACADE, a practical new paradigm that enhances the
confidentiality, integrity, and availability of the files stored in a public cloud.  
\FACADE
attained these objectives by utilizing a fast and light-weight XOR secret
sharing scheme and leveraging the power of publicly available cloud platforms
in a transparent manner to essentially build a ``cloud of clouds''.  Using XOR
secret sharing, \FACADE generates three secret shares of the user's data file,
whereby learning just one of these shares reveals no meaningful information
about the file and at least two secrets are required to build the original
file. Such an approach enhances the availability of data in case one of the cloud
providers becomes inaccessible (or temporarily congested). \FACADE also guarantees the
integrity of the data, as any modification of at most two of the secret shares
will impact the reconstruction of the file (the reconstructed file will be
garbage). \FACADE can provide all these properties, while offering the user
experience similar to that of password authentication as users can be given
access to their cloud-stored files with just a single master password, which 
locks the access tokens used to authenticate to each of the
cloud service providers. Our design, implementation, and evaluation of \FACADE
show that it is efficient enough to be readily employed by the users to
improve the confidentiality, availability, and integrity of their sensitive
data.

{\footnotesize \bibliographystyle{ACM-Reference-Format}
	\bibliography{bibliography}}

\section{Appendix}
\label{sec:appedix}

\noindent \begin{minipage}[t]{.45\textwidth}
\begin{algorithm}[H] 
\scriptsize
\caption{PASSAT Secret Sharing}
\label{secretshare2}
\begin{algorithmic}[1]

\Function{Split}{$file ~ f$}
\State{\textbf{if} $size(file) > threshold_t$ \textbf{then}}
\State{\indent {$\{f_1\dots,f_n\}$}  $\leftarrow$ $f/t$}
\State{\indent {$ tag $ $ t_i\mapsto f_i |  i=1,2,\dots,n $}}
\State \Return {$(f_i,t_i) = d_i$}
\EndFunction

\Function{Secret Share}{$d$}			
\State {$s_0 \leftarrow  0^d, s_1 || \dots || s_{n_{p}}-1  \leftarrow s$}
\For{$i \leftarrow 0$ to $k - 2 $}                  
\For{$j \leftarrow 0$  to $n_p -1 $}
\State {$r_i^j \leftarrow GEN (\{0, 1\}^d) $}

\EndFor
\EndFor{(discard $r_{n_{p}}^0$ - 1)}

\For{$i \leftarrow 0$ to $n - 1 $}                  
\For{$j \leftarrow 0$ to $n_p -2 $}
\State {$S_{(i,j)} \leftarrow  (\oplus_{h=0}^{k-2}  r_{h.i+j}^h) \oplus s_j-i$}

\EndFor
\State{$S_i \leftarrow S_{(i,0)} || \dots || S_{(i,n_{p}-2)}$}
\EndFor	
\State \Return {$\{S_0\dots,S_{n-1}\}$}
\EndFunction

\Function{auth}	{}

\State {$Lastpass.retrive $ $ \{tk_1,tk_2,tk_3\} $}

\For{$i =1$ to $n$}
\State {$Cloud $ $ c_i \leftarrow S_i$}

\EndFor
\EndFunction		

\end{algorithmic}
\end{algorithm}
\end{minipage} \hfill
\noindent \begin{minipage}[t]{.45\textwidth}
\begin{algorithm}[H] 
	\scriptsize
\caption{PASSAT Reconstruction}
\label{reconstruct2}
\begin{algorithmic}[1]

%

\Function{auth}	{}
\State {$Lastpass.retrive $ $ \{tk_1,tk_2,tk_3\} $}

\For{$ i =1$ to $n$}
\State {$S_i \leftarrow c_i$}

\EndFor
\EndFunction		

\Function{Secret Reconstruct}{$d$}			
\For{$i \leftarrow 0$ to $k - 1$}
\State{$ S_{(S_{t_{i}},0)} || \dots || S_{(i,n_p-2) \leftarrow S_{t_{i}}}$}
\EndFor
\State{$ S \leftarrow (S_{(S_{t_{0}},0)},\dots, S_{(S_{t_{0}},n_p-2)},$\newline$\dots,
S_{(S_{t_{k-1}},0)},\dots,S_{(S_{t_{k-1}},{n_p-2})})^T$}

\State {$ M \leftarrow MAT (t_0,\dots,t_k -1)$}
\State {$ ((s_1,\dots,s_{n_p} -1))^T \leftarrow M.S$}
\State {$ s \leftarrow s_1 ||\dots || s_{n_p}-1$}
\State \Return {$s$}

\EndFunction

\Function{merge}{$secret $ $ s$}
\State{\textbf{if} $t_n \exists $ $ s $ \textbf{then}}				
\For{$s, t_n $}

\State{$ f \leftarrow Concatenate(s_{t_{1}},\dots,s_{t_{n}}) $}
\State \Return {$f$}
\EndFor
\EndFunction	

\end{algorithmic}
\end{algorithm}

\end{minipage}

\end{document}